\documentclass{paper}

\usepackage[style=alphabetic, sorting=nyt, natbib, backend=biber, backref=true, backrefstyle=none, maxnames=5, maxalphanames=5]{biblatex}
\usepackage[bitstream-charter]{mathdesign}
\usepackage[T1]{fontenc}
\usepackage{multicol}
\usepackage{subcaption}
\usepackage{enumitem}
\usepackage{lipsum}

\makeatletter
\def\myinline#1#2{%
\ifx\@footnotetext\TX@trial@ftn
\detokenize{#2}%
\else
\mintinline{#1}{#2}%
\fi}

\makeatother

\newcommand{\onlinecompf}[3]{\left\{ #1, #2, #3 \right\}}
\newcommand{\lambdacompf}[2]{\onlinecompf{O\left(n \lambda_k(n)\right)}{#1}{#2}}
\newcommand{\loglambdacompf}[2]{\onlinecompf{O\left(n \log \lambda_k(n)\right)}{#1}{#2}}

\newcommand{\code}[1]{\left\langle #1 \right\rangle}

\title{Tree Path Minimum Query Oracle via Boruvka Trees}

\author{Tianqi Yang \vspace{0.7ex} \\ \textit{Institute for Interdisciplinary Information Sciences} \\ \textit{Tsinghua University} \\ \textit{Beijing, China} \\ \href{mailto:yantq19@mails.tsinghua.edu.cn}{\nolinkurl{yangtq19@mails.tsinghua.edu.cn}}}

\begin{document}
    
    \maketitle
    
    \begin{abstract}
        Tree path minimum query problem is a fundamental problem while processing trees, and is used widely in minimum spanning tree verification and randomized minimum spanning tree algorithms. In this paper, we study the possibility of building an oracle in advance, which is able to answer the queries efficiently. We present an algorithm based on Boruvka trees. Our algorithm is the first to achieve a near-optimal bound on query time, while matching the currently optimal trade-off between construction time and the number of comparisons required at query. Particularly, in order to answer each query within $2k$ comparisons, our algorithm requires $O(n \log \lambda_k(n))$ time and space to construct the oracle, and the oracle can answer queries in $O(k + \log \lambda_k(n))$ time. Here $\lambda_k(n)$ is the inverse of the Ackermann function along the $k$-th column. This algorithm not only is simpler than the previous ones, but also gives a completely different method of solving this problem.
    \end{abstract}
    
    \section{Introduction} \label{sec:intro}
        The tree path minimum query problem is to find the minimum weight along the simple path from one node to another on a tree. It contributes notably to the problem of minimum spanning tree verification \citep{komlos1985linear,king1997simpler}, which has been shown to imply efficient randomized minimum spanning tree algorithm \citep{karger1995randomized}. To verify whether a spanning tree is minimum, we only need to check, for each edge not on the spanning tree, the maximum weight along the simple path between the two endpoints of the edge on the spanning tree. This only requires an `offline' solution to path minimum query, for which we means that all queries are given as a large batch, and the algorithm is allowed to process them simultaneously. For this reason, this `offline' version of this problem has been studied extensively in the literature. The first linear algorithm is presented by \citet{komlos1985linear}, but the algorithm only achieves linear in the number of comparisons used. The algorithm itself is far from linear. \citet{dixon1992verification} come out with the first \emph{truly} linear-time algorithm, but this initial proposal is hard to implement. Motivated by this, several simplifications are later made based upon the Komlos's algorithm, including King's algorithm based on the Boruvka trees \citep{king1997simpler}, and Hagerup's algorithm based on set theory \citep{hagerup2009even}.
        
        However, if the queries come in \emph{one-by-one}, and we are required to prepare an oracle to answer these queries online, then all the above algorithms will no longer work. As a special case, the famous range minimum query problem adopts a linear-time solution in word RAM which can answer the queries in constant time \citep{bender2000lca}. However, this algorithm builds upon the Cartesian tree of the sequence, which cannot be built linearly on a general tree (as we will discuss in Section \ref{sec:pre:cartesian}). In fact, answering these queries online is intrinsically difficult, in the sense that a lower bound is known that $\Omega(n \log \lambda_k(n))$ pre-processing time is necessary to answer queries within $k$ comparisons \citep{pettie2006inverse}, where $\lambda_k(n)$ is the inverse of the Ackermann function along the $k$-th column. We will formally define this function in Section \ref{sec:pre:ackermann}.
        
        Apart from this result, algorithms are known to nearly match this lower bound. Built upon the Yao's algorithm for partial sums in a linear list \citep{yao1982space}, \citet{alon1987optimal} proposed an algorithm that builds an oracle in $O(n \lambda_k(n))$ time and space to answer the queries within $4k$ comparisons. Another approach is presented by \citet{chazelle1987computing} in the same year with the same preprocessing time and $2k+O(1)$ query comparisons\footnote{In fact, these two algorithms work for a more generalized setting called semi-group queries, where we query the sum of all weights on a path in a semi-group.}. Based on the two algorithms above, \citet{pettie2006inverse} claimed an oracle which can answer each query in $2k$ comparisons, and can be constructed in $O(n \log \lambda_k(n))$ time and space. This is the best known result for this problem, but it gives no guarantee on the query time of the oracle.
        
        In this paper, we present the first algorithm which, while keeping the best known preprocessing time and query comparisons needed, gives a near-optimal bound on the query time required in word RAM. Particularly, our algorithm constructs an oracle in $O(n \log \lambda_k(n))$ time and space, which is able to answer the queries within $2k$ comparisons, and $O(k + \log \lambda_k(n))$ time. If $O(k)$ time is intended, then we either need to loosen the preprocessing time to $O(n \lambda_k(n))$, or the number of comparisons needed to $2k + 2$. Moreover, our algorithm is, in our eyes, much simpler than the previous ones, in both the algorithm itself and the analysis.
        
        \subsection{Intuition}
            Our algorithm is based on the Boruvka trees introduced by \citet{king1997simpler}. Briefly speaking, Boruvka tree is the structure built during Boruvka's maximum spanning tree algorithm. It has the beautiful property of preserving path minimum query: the minimum weight on the path between any two nodes in a tree is exactly equal to the minimum weight on the path between the two corresponding nodes in its Boruvka tree.
            
            Boruvka trees also have some additional useful properties, making it easier for us to handle path minimum queries on it than general trees. These properties are similar to full binary trees: all leaves of a Boruvka tree have the same depth, and we will show later in Section \ref{sec:alg:balanced-boruvka} that we can further make the number of children of each internal node between $2$ and some constant $c$. These properties give us an upper bound on the number of vertices with small depth. Hence we can preprocess all vertices with small depth (depth smaller than a threshold $s$) via a trivial algorithm, which is affordable since the number of vertices is not too large. Then we divide the rest of tree into many smaller parts, each of which can be recursively solved. For any query, we can then divide the path into three segments, with the middle one having depth less than $s$, and the other two in some smaller subtrees. By carefully setting the threshold $s$, we can construct, in $O(n \lambda_1(n))$ time, an oracle which can answer the queries with $2 \lambda_1(n)$ comparisons. With a simple trick at preprocessing, the number of comparisons can be reduced down to $2$. Moreover, the queries can be answered in constant time.
            
            Then we can repeat this process, replacing the simple algorithm to preprocess the small depth cases, with the above non-trivial algorithm with $O(n \lambda_1(n))$ preprocessing time and $2$ query comparisons. By re-choosing the thresholds, we can construct an oracle in $O(n \lambda_2(n))$ time to answer the queries with $4$ comparisons. By doing this $k$ steps, we get an oracle to answer the queries with $2k$ comparisons, which can be constructed in $O(n \lambda_k(n))$ time. Also, the query time is bounded by $O(k)$. By analyzing the bottleneck on constructing the oracle, preprocessing time can be further reduced to $O(n \log \lambda_k(n))$ by sacrificing either $2$ additional queries, or $O(\log \lambda_k(n))$ query time.
    
    \section{Preliminaries} \label{sec:pre}
        In this paper, we mainly focus on edge-weighted trees $T = \code{V, E, W}$, where $V$ is the set of vertices, $E$ is the set of edges, and $W$ corresponds to the weights of the tree. We usually use $n$ to denote the number of vertices in the tree. For convenience, we adopt the notation of using $fa_u$ to represent the father of $u$ and $ch_u$ to represent the \emph{children set} of $u$. For rooted trees, we can further define the \emph{depth} of a node to be the number of vertices on the simple path from the root to it, denoted by $dep_u$. In particular, the root has depth $1$. We call the maximum depth of all nodes the \emph{height} of the tree. We can also define the lower common ancestor of two nodes $u$ and $v$ in the normal way, denoted by $LCA(u,v)$.
        
        The problem we consider can be formalized as the following: for a fixed tree $T$, construct a data structure, which on queries of the form $(u,v)$, answer the minimum weight of all edges on the unique simple path from $u$ to $v$. We note here that this edge-weighted version is in fact equivalent to the node-weighted variant, with a linear pre-processing time overhead, since we can reduce the latter one to the former by taking the weight of each edge to be the smaller one between the two endpoints, and conversely by adding dummy nodes on the edges.
        
        What is critical in this problem is that the queries come in one-by-one, and the data structure we construct needs to answer them \emph{online}. For the sake of simpler notation, if an algorithm requires $T_1(n)$ preprocessing time, and can then answer any query within $T_2(n)$ comparisons, and $T_3(n)$ additional time in word RAM model, then we denote the complexity by $\onlinecompf{T_1(n)}{T_2(n)}{T_3(n)}$. For example, the algorithm we are going to propose has complexity $\loglambdacompf{2k}{O(k+\log \lambda_k(n))}$.
        
        \subsection{Boruvka trees} \label{sec:pre-boruvka}
            The key component we utilize is the Boruvka trees, introduced by \citet{king1997simpler}, which is the hierarchical structure constructed during the process of Boruvka's algorithm. Initially, a leaf is constructed in Boruvka tree for each node in the original tree $T$. Then we do one iteration of Boruvka algorithm, which marks out the minimum edge incident to each node, and shrink the connected components formed by marked edges into hypernodes. For each shrunk connected component, say $C$, we create a node $v_C$ for it in the Boruvka tree, then for any $v \in C$, we set the father of $v$ in the Boruvka tree to be $v_C$, with the edge weight equal to the incident edge selected by $v$ in this iteration. Then we use the $v_C$'s to represent the hypernodes created in this iteration, and repeat the iterations until only one hypernode containing the whole tree left. Since the time complexity of Boruvka algorithm on trees is $O(n)$, we can construct the Boruvka tree in linear time. Without further clarification, we consider the Boruvka’s algorithm for \emph{maximum spanning trees} in the rest of this paper.
            
            For convenience, for a tree $T$, we denote the corresponding Boruvka tree (given by the \emph{maximum spanning tree} variant of Boruvka's algorithm) by $B(T)$. For each node $u$ in $T$, we denote the corresponding leaf in $B(T)$ by $B(u)$. The key property of such a Boruvka tree is captured in \citep{king1997simpler}, which relates the path minimum query problems on $T$ and $B(T)$.
            
            \begin{lemma}[Theorem 1 of \citep{king1997simpler}] \label{lmm:boruvka-eq}
                Let $T$ be any edge-weighted tree, and $B(T)$ be its corresponding Boruvka tree (given by the \emph{maximum spanning tree} variant of Boruvka's algorithm), then for any pair of vertices $u, v$ in $T$, the minimum weight on the path between $u$ and $v$ in $T$ is equal to the minimum weight on the path between $B(u)$ and $B(v)$ in $B(T)$.
            \end{lemma}
            
            This lemma tells us that in order to answer path minimum query on $T$, we only need to get the minimum weight on the corresponding path on $B(T)$. This allows us to reduce path minimum query on \emph{general trees} to \emph{Boruvka trees}. In the meantime, Boruvka trees have two wonderful properties:
            \begin{itemize}
                \item all leaves in a Boruvka tree have the same depth,
                \item all internal nodes in a Boruvka tree have at least two children.
            \end{itemize}
            These properties can both be shown directly from the definition, so detailed proof is omitted. The point is, they allows us to bound the number of vertices near the root.
            \begin{lemma} \label{lmm:boruvka-subtree}
                For a Boruvka tree $B$ of height $h$, the number of nodes with depth not larger than $k$ is at most $\frac{n}{2^{h-k}}$.
            \end{lemma}
            
            \begin{proof}
                Since each internal node has at least two children, for a node $u$ at depth $k$, the subtree rooted at $u$ has size at least $2^{h-k+1}-1$. Also, similar to a full binary tree, the number of nodes with depth $< k$ is smaller than those with depth exactly $k$. Suppose that the number of vertices with depth less than $k$ is $a$, and the number of vertices with depth exactly $k$ is $b$, then we should have $a < b$, and the number of vertices with depth not smaller than $k$ is at least $(2^{h-k+1}-1)b$. Hence, the maximum number of vertices with depth not larger than $k$ is given by the following linear program:
                \begin{align*}
                    \max.\ \ & a + b \\
                    s.t.\ \ & a < b \\
                    & a + (2^{h-k+1}-1)b \le n.
                \end{align*}
                Solving this completes the proof.
            \end{proof}
            
            As a simple corollary, we can further imply that any Boruvka tree has height $O(\log n)$.
            
        \subsection{Cartesian trees} \label{sec:pre:cartesian}
            Another useful data structure is the Cartesian tree introduced by \citet{vuillemin1980unifying} to capture the order of elements in a permutation. This idea can naturally be generalized to trees. For simplicity, in the rest of this sub-section, we consider node-weighted trees. However, as we have noted earlier, this can be easily translated into an edge-weighted version.
            
            For an unrooted, node-weighted tree $T$, we can define the corresponding Cartesian tree $C(T)$ as a rooted tree satisfying the following three properties:
            \begin{itemize}
                \item the nodes of $C(T)$ has a bijective correspondence with the nodes in $T$,
                \item for each subtree of $C(T)$, all nodes in it form a connected subgraph in $T$,
                \item $C(T)$ satisfies the heap property, i.e., the weight of each node cannot be greater than the weight of its children.
            \end{itemize}
            
            If the weights on the vertices are pair-wise different, then the Cartesian tree is unique, since we can only recursively select the vertex with smallest weight to be the root in order to satisfy the heap property.
            
            Assuming that we already know the order of all weights (for example, the weights are given as a permutation), Cartesian tree of a tree can be contructed in linear time \citep{chazelle1987computing} using the linear-time union-find on trees given by \citet{gabow1985linear}.

            With the Cartesian trees, we can then easily answer tree path minimum query, since the node with the minimum weight on path between $u$ and $v$ is exactly the lowest common ancestor of $u$ and $v$ in $C(T)$. Lowest common ancestor can be answered in constant time with linear-time prepocessing \citep{bender2000lca}, so we get a simple $\onlinecompf{O(n \log n)}{0}{O(1)}$ algorithm for tree path minimum query, with the bottleneck at preprocessing being sorting all the weights while building Cartesian trees.
            
            In fact, this sorting is unavoidable in building Cartesian trees for trees. For any sequence $a_1, a_2, \dots, a_n$, we can construct a tree of size $n+1$ in the following manner: the $i$-th vertex (for $1 \le i \le n$) has weight $a_i$, and has an edge to the $(n+1)$-th vertex. The $(n+1)$-th vertex has a weight larger than any of the other $n$ elements. Then building the Cartesian tree of this tree will solve sorting, hence gives us a $\Omega(n \log n)$ lower bound for building Cartesian trees.
            
        \subsection{Ackermann function} \label{sec:pre:ackermann}
            The Ackermann function \citep{ackermann1928hilbertschen} is used widely in complexity analysis. Here we give a slightly modified definition of this function and its inverses:
            \begin{definition}[Ackermann Function] \label{def:ackermann}
                \begin{equation*}
                    A(m,n) = \begin{cases}
                        2^n & m = 0 \\
                        A(m-1,1) & m \ge 1, n = 0 \\
                        A(m-1,A(m,n-1)) & m \ge 1, n \ge 1
                    \end{cases}.
                \end{equation*}
            \end{definition}
            
            \begin{definition}[Inverse of Ackermann Function along Columns] \label{def:ackermann-inverse-column}
                \begin{equation*}
                    \alpha (m,n) = \min \left\{ i \ge 1: A \left(i, \left\lfloor \frac{m}{n} \right\rfloor\right) \ge n \right\}.
                \end{equation*}
                For simplicity, we further use $\alpha(n)$ as a shorthand for $\alpha(n,n)$.
            \end{definition}
            
            \begin{definition}[Inverse of Ackermann Function along Rows] \label{def:ackermann-inverse-row}
                \begin{equation*}
                    \lambda_k (n) = \min \left\{ j \ge 1: A(k,j) \ge n \right\}.
                \end{equation*}
            \end{definition}
            
            From the definition, it can be seen immediately that
            \begin{equation} \label{eq:ackermann-inverse-eq}
                \lambda_{\alpha(n)}(n) = 1.
            \end{equation}
            
    \section{Main algorithm on Boruvka trees}
        \subsection{Balanced Boruvka trees} \label{sec:alg:balanced-boruvka}
            We first present a modified version of the Boruvka's algorithm. Recall that for each round, the algorithm will pick, for each vertex, the edge incident to it with the maximum weight (again we emphasize that we are finding the maximum spanning tree), and shrink all the connected components formed by picked edges to a hyper-node. In our modified version, we will take a constant $c$, and at the beginning of each round, first make the degree of each node in the tree not larger than $c$ by splitting nodes with large degrees. This will involve at most $\frac{n}{c-2}$ additional nodes. Then after picking the edges with maximum weights, we will repeatedly drop the middle edge for any simple path formed by picked edges of length $3$. This will make the diameter of all the remaining connected components not greater than $2$, while keeping the assertion that there is a least one picked edge adjacent to each node. Then we shrink the connected components in the normal way.
            
            \begin{algorithm}[htb]
                \caption{A modified Boruvka's algorithm}
                \label{alg:modified-boruvka}
                \begin{algorithmic}
                    \While{The graph has more than $1$ node}
                        \While{There is a node $u$ with degree larger than $c$}
                            \State Split $u$ into two nodes both with weights equal to $u$
                        \EndWhile
                        \State For each vertex, mark the edge incident to it with maximum weight
                        \While{There is a path $u-a-b-t$ of length $3$ formed by marked edges}
                            \State Unmark the edge $(a,b)$
                        \EndWhile
                        \State Shrink all connected components formed by marked edges into hypernodes
                    \EndWhile
                \end{algorithmic}
            \end{algorithm}
            
            The reason why we apply this modification is that this will ensure that the size of each shrunk connected component is not larger than $c+1$, since its diameter is not greater than $2$, and the degree of each node is not greater than $c$. In the meantime, since we still guarantee that each vertex has at least one marked edge incident to it, the number of vertices will reduce to half after the shrinkage. Hence the size of the shrunk tree after each round is bounded by
            \begin{equation*}
                \frac{1}{2} \left( n + \frac{n}{c-2} \right) = \frac{c-1}{2c-4} n.
            \end{equation*}
            By picking $c \ge 4$, we have $\frac{c-1}{2c-4} < 1$, so the algorithm is still linear, meaning that the size of the corresponding Boruvka tree is linear as well.
            
            The tree generated by this modified Boruvka's algorithm have all the properties in Section \ref{sec:pre-boruvka}. In addition, the number of children of each node can never be greater than $c+1$ (i.e., $\lvert ch_u \rvert \le c+1$). Thus the degree of each node is bounded by a constant, which give us a tight bound that $h = \Theta(\log n)$, where $h$ is the height of the tree and $n$ is the number of vertices in it. We call it the \emph{balanced Boruvka tree} of the original tree $T$, denoted by $B'(T)$.
            
        \subsection{Basic algorithm} \label{sec:alg:basic}
            By lemma \ref{lmm:boruvka-eq}, we can translate the problem on $T$ to the equivalent problem on $B'(T)$ in linear time, so we now only consider the queires on $B'(T)$. Suppose that the tree has $n$ vertices and height $h$, then $h = \Theta(\log n)$. In the balanced Boruvka tree $B'(T)$, the size of $ch_u$ for all nodes $u$ is bounded by a constant $c' = c + 1$.
            
            We now present two simple ways of getting an $\onlinecompf{O(nh)}{0}{O(1)}$ algorithm on such a balanced Boruvka tree $B'(T)$. The first one is by applying the Cartesian tree. As discussed in Section \ref{sec:pre:cartesian}, we can achieve $\onlinecompf{O(n \log n)}{0}{O(1)}$ by constructing the corresponding Cartesian tree. Since $h = \Theta(\log n)$, $O(n \log n)$ is equivalent to $O(nh)$.
            
            The other method is more straight-forward. We maintain at each node $u$, the order of the answers from $u$ to every node $v$ in the subtree of $u$ in $B'(T)$. By applying merge sort on this tree structure, the time complexity to process all the nodes is (here we abuse the notation that $ch_v$ and $dep_v$ are defined on $B'(T)$, and $B'(T)(u)$ represents the subtree in $B'(T)$ rooted at $u$)
            \begin{equation*}
                \sum_{v \in V} \lvert ch_v \rvert \cdot \lvert B'(T)(v) \rvert \le \sum_{v \in V} c' \cdot \lvert B'(T)(v) \rvert = c' \sum_{v \in V} dep_v \le c'nh = O(nh).
            \end{equation*}
            Querying can be answered by checking the order of the result from $u$ to $LCA(u,v)$ and the result from $v$ to $LCA(u,v)$. This gives us an $\onlinecompf{O(nh)}{0}{O(1)}$ algorithm.
            
        \subsection{Recursion to speed up preprocessing} \label{sec:alg:recursion-1}
            Our intuition is somehow reducing either $n$ or $h$ to make the algorithm affordable. Let us set a threshold $s$, and solve the cases where both ends of the query have $dep \le s$ with the basic algorithm. By Lemma \ref{lmm:boruvka-subtree}, the number of such nodes is $O \displaystyle \left(\frac{n}{2^{h-s}}\right)$. Setting $s = h - \log h$, the complexity of preprocessing becomes
            $O \displaystyle \left( \frac{n}{2^{\log h}} \cdot h \right) = O(n)$, which is efficient. We then solve the remaining cases recursively.
            
            Formalizing the idea above, we set $m$ thresholds $s_1, s_2, \dots, s_m$, where $s_i = h - \log^{(i)} h$, with
            \begin{equation*}
                \log^{(i)} h = \begin{cases}
                    h, & i = 0 \\
                    \log \log^{(i-1)} h, & i > 0
                \end{cases}.
            \end{equation*}
            Then $m = \log^* h = \lambda_1(h)$. For simplicity, we assume that $s_0 = 0$. For each layer $i$, we process all the nodes within the depth range from $s_{i-1}$ to $s_i$ using the basic algorithm. Since the number of nodes in each layer is bounded by Lemma \ref{lmm:boruvka-subtree}, the complexity of each layer is
            \begin{equation*}
                O \left( \frac{n}{2^{\log^{(i)} h}} \cdot \log^{(i-1)} h \right) = O(n).
            \end{equation*}
            Since there are $m = \lambda_1(h)$ layers, the total time complexity of preprocessing is $O\left(n \lambda_1(h)\right)$. In addition, for each node $u$ and layer $i$, suppose that the ancestor of $u$ with depth exactly $s_i$ is $p_{u,i}$, then we compute the minimum weight on the path from $u$ to $p_{u,i}$ in advance. This can also be done in $O\left(n \lambda_1(h)\right)$ time and space.
            
            To answer a query $(u,v)$ with the above information, we first find out their lowest common ancestor $l = LCA(u,v)$. Suppose that $l$ is in layer $i$, then we split the full path into three segments: the part in layer $i$, and the parts from $u$ and $v$ to layer $i$ respectively. The answer of all these segments can be directly found in the preprocessed information. Thus we only need $2$ comparisons to find the minimum one among them. In this way, we obtained a solution with time and space complexity $\onlinecompf{O \left( n \lambda_1(h) \right)}{2}{O(1)}$
            
        \subsection{Recursion of recursions} \label{sec:alg:recursion-2}
            We call the algorithm described in the previous sub-section the first step, and $f_1(n,h)$ to be its preprocessing time and space complexity. Then $f_1(n,h) = O \left( n \lambda_1(h) \right)$. Our intuition is to repeat this process with different thresholds. For $k \ge 2$, we suppose that the complexity of step $k-1$ is $f_{k-1}(n,h) = O \left( n \lambda_{k-1}(h) \right)$, we now consider the $k$-th step.
            
            We set the thresholds $s_i$ to be $h - \lambda_{k-1}^{(i)} (h)$, with
            \begin{equation*}
                \lambda_{k-1}^{(i)}(h) = \begin{cases}
                    h, & i = 0 \\
                    \lambda_{k-1}(\lambda_{k-1}^{(i-1)}(h)), & i > 0
                \end{cases},
            \end{equation*}
            then the number of layers is $m = \lambda_k (h)$. Now we process each layer in the same way as Section \ref{sec:alg:recursion-1}, except that we handle the nodes with depth exactly $s_i$ individually. Specifically, we preprocess the nodes within the depth range from $s_{i-1}+1$ to $s_i-1$ using the algorithm at step $k-1$ which preprocessing time complexity $f_{k-1}(n,h)$, and the answers from each node to their ancestors with depth exactly $s_i$ or $s_i-1$. By dealing with the boundary of the layers carefully at query time, we can keep the query complexity to be the same. In this way, the number of nodes to be preprocessed at layer $i$ using the algorithm of step $k-1$ will be bounded by $\displaystyle \frac{n}{2^{h-s_i+1}}$.
            
            Suppose that the number of nodes in the depth range from $s_{i-1}+1$ to $s_i-1$ is $n_i$, then they meet the following requirement:
            \begin{equation*}
                n_1 + n_2 + \dots + n_i \le \displaystyle \frac{n}{2^{\lambda_{k-1}^{(i)}(h)+1}},\ \forall\ i
            \end{equation*}
            Taking $i = m$ gives us $n_1 + n_2 + \dots + n_m \le \frac{n}{2}$. The preprocessing time complexity of this step is
            \begin{equation} \label{eq:preprocess}
                f_k(n,h) = \sum_{i=1}^m f_{k-1} \left( n_i, \lambda_{k-1}^{(i-1)} (h) \right) + O \left( n \lambda_k (h) \right).
            \end{equation}
            By solving this equation, we can get a bound for the preprocessing time complexity of step $k$.
            
            \begin{theorem} \label{thm:preprocess}
                For any $k$ which can depend on $n$, $f_k(n,h) = O \left( n \lambda_k (h) \right)$.
            \end{theorem}
            
            \begin{proof}
                The proof is by induction on $k$. The statements is trivial while $k = 1$.
                
                For any $k \ge 2$, suppose that $f_{k-1}(n,h) \le c_1 n + c_2 n \lambda_k (h)$ for some constant $c_1, c_2$. Then,
                \begingroup
                    \allowdisplaybreaks
                    \begin{align*}
                        f_k(n,h) & = \sum_{i=1}^m f_{k-1} \left( n_i, \lambda_{k-1}^{(i-1)} (h) \right) + O \left( n \lambda_k (h) \right) \\
                        & \le \sum_{i=1}^m \left( c_1 n_i + c_2 n_i \lambda_{k-1}^{(i)} (h) \right) + O \left( n \lambda_k (h) \right) \\
                        & = c_1 \sum_{i=1}^m n_i + c_2 \sum_{i=1}^m n_i \lambda_{k-1}^{(i)} (h) + O \left( n \lambda_k (h) \right) \\
                        & \le c_1 \frac{n}{2} + c_2 \sum_{i=1}^m \frac{n}{2^{\lambda_{k-1}^{(i)}(h)+1}} \lambda_{k-1}^{(i)}(h) + O \left( n \lambda_k (h) \right) \\
                        & \le c_1 \frac{n}{2} + c_2 \frac{n}{2} \sum_{i=1}^n \frac{i}{2^i} + O \left( n \lambda_k (h) \right) \\
                        & = \left( \frac{c_1}{2} + c_2 \right) n + O \left( n \lambda_k (h) \right).
                    \end{align*}
                \endgroup
                
                From the result, we can see that the constant $c_1$ will not increase with $k$, so we can discard the first term, and get $f_k(n,h) = O(n \lambda_k(h))$.
            \end{proof}
            
            Note that $h = \Theta(\log n)$, so this also guarantees that the preprocessing time is in $O \left( n \lambda_k(n) \right)$.
            
            At each step, the query path from $u$ to $v$ will be split into three parts, so $2$ additional comparisons are needed at each step. Therefore, the number of comparisons needed for each query is $2k$, with query complexity $O(k)$. Combining with the basic algorithm described in Section \ref{sec:alg:basic}, which can be considered as the $k=0$ case, we obtain an $\lambdacompf{2k}{O(k)}$ algorithm.
            
            By Theorem \ref{thm:preprocess}, this complexity holds even if $k$ is relevant to $n$. Taking $k = \alpha(n)$, combining with Equation \eqref{eq:ackermann-inverse-eq}, the algorithm will become $\onlinecompf{O(n)}{\alpha(n)}{O\left(\alpha(n)\right)}$.
            
        \subsection{Further improvements} \label{sec:alg:improvement}
            The bottleneck of the current algorithm is the $O \left( n \lambda_k (h) \right)$ term in Equation \eqref{eq:preprocess}, which is the time needed to preprocess the answers from each node to their ancestors on the layer borders at the last step. This part is essentially a leaf-to-ancestor query on a tree with $n$ nodes and height $O \left( \lambda_k (h) \right)$. The Komlos's algorithm for minimum spanning tree verification \citep{komlos1985linear} provides us an $\onlinecompf{O(n \log h)}{0}{O(\log h)}$ algorithm. Komlos's algorithm pre-computes the answer of each node by traversing the tree once with a stack and we can use a persistent balanced binary search tree (e.g., treap or red-black tree) to maintain the stack. This implies an $\loglambdacompf{2k}{O\left(k + \log \lambda_k (n)\right)}$ algorithm for our problem, which is faster to preprocess while requiring more query time, although the number of comparisons to answer a query remains the same.
            
            This is in fact a trade-off among the preprocessing time, the number of comparisons needed to answer a query and the query time. Indeed, an $\loglambdacompf{2k+2}{O(k)}$ solution can also be obtained by applying long path decomposition at this last step.
            
    \section{Open problems}
        As discussed above, our result suffers a trade-off among the preprocessing time, the number of comparisons needed to answer a query, and the additional query time needed in word RAM model. One direct open problem is to find a $\loglambdacompf{2k}{O(k)}$ algorithm. To solve this problem, a $\onlinecompf{O(n \log h)}{0}{O(1)}$ solution for leaf-to-ancestor queries would be sufficient. This result is appealing since it gives an algorithm which matches the currently best algorithm on the number of comparisons, and is also optimally fast practically in word RAM.
        
        Although the performance of our algorithm matches previous works, there is still a gap between the complexity of our algorithm and the proven lower bound for this problem in \citep{pettie2006inverse}. Closing this gap would also be a very interesting future direction. Indeed, our algorithm provides a novel way of handling this problem, compared with previous ones based on \citep{yao1982space}. There might be chances to combine the ideas together to obtain a better solution.
        
    \paragraph{Acknowledgement.}
        I am grateful to Zhiyuan Fan, Jiatu Li and Yiding Zhang for many useful discussions throughout this work.

    \printbibliography
    
\end{document}